# Control Strategy for Virtual Synchronous Generator Based on Y-Z Source Inverter in Islanded Grids

Shirin Besati[1,3], Ali Mosallanejad[2], and Madhav Manjrekar[1,3]
[1]Department of Electrical and Computer Engineering, University of North Carolina at Charlotte, USA
[2]Shahid Beheshti University, Tehran, Iran
[3]Energy Production and Infrastructure Center, University of North Carolina at Charlotte, USA
sbesati@uncc.edu, a_mosallanejad@sbu.ac.ir, mmanjrek@uncc.edu

*Abstract*— The Virtual Synchronous Machine (VSM) concept stands as a strategy for the seamless integration of renewable energy sources into the grid. In scenarios where symmetrical and sinusoidal operating conditions are lacking, VSMs demonstrate multifaceted capabilities. They can operate as power sources especially in islanded-systems, adeptly address harmonic distortions and imbalances, offer load power compensation, and elevate voltage quality at the grid. Even so, dead time variations, discontinuous input current, and non-single-power-stage configuration of conventional voltage/current source inverters and also, load variation can result in VSMs behaving as harmonic and imbalance absorbers. This paper utilizes a specific Y-Source Inverter (YZSI) for the first time as a solution to these challenges. Through integrating this innovative configuration with an exceptionally adaptable voltage gain and an enhanced multi-loop control strategy into an isolated power system accommodating diverse load profiles, it not only achieves high efficiency in power transfer but also enhances overall system performance quality. Furthermore, it contributes to extending the lifespan and preserving the quality of the input source. The controller is designed to regulate system frequency, voltage, and manage voltage and current for YZSI, delivering impressive performance even in non-ideal grid conditions. MATLAB/Simulink confirms the efficacy of the proposed YZSI VSG control strategy with a high accuracy achieved through a comprehensive examination of the internal component parameters.

I. INTRODUCTION

Nowadays, contamination has been under specific consideration. One reasonable way is utilizing renewable energy sources in the grid and decentralizing grids [1-3]. To actualize this purpose, the necessity for power electronic converters and specific controllers has emerged. Voltage/ current inverter VSG layouts are widely utilized in grid compensations as a grid-following aspect and in islanded systems independently as grid-forming aspect [4]. Virtual synchronous machine (VSM) control is an approach to provide power electronic converters with an inertia and a damping factor virtually to imitate the behavior of a synchronous machine [5-7]. Beck and Hesse in 2007 introduced "VISMA" as the first suggestion of a VSM [8]. Afterward, Countless methodologies have appeared for integrating virtual characteristics of a synchronous machine into power electronic converters, that each presents Explicit purpose according to their Obligations [9]. In this context, only two conventional power electronic inverters, namely Voltage Source Inverters (VSI) and Current Source Inverters (ISI), are generally served to attach the controller to renewable energy sources. So, VSI/CSI VSGs along with their benefits, can simultaneously generate significant challenges within grid systems, notably in grid-forming conditions, because they must provide secure management in power generation, distribution, and compensation [10-16]. In this section, a concise overview of these matters is presented below:

- The switching bridge of the inverters can make harmonics and other power quality issues into the grid. Accordingly, the inverters may need to provide a reactive compensator to uphold grid voltage stability;
- In islanded networks, because of irregular renewable energy resources and variations in loads sustaining stable voltage and frequency is challenging, therefore, VSIs and CSIs are required to adjust these parameters to guarantee adaptability to these situations;
- After an outage or during a variation in load demands (or injection of an inductive load), when the power supply needs to reconnect to the grid or adapt itself to the new circumstance, the inverters must synchronize both output voltage magnitude and frequency with the grid to avoid power quality issues, voltage/current unbalancing, and equipment damage during reconnection.

Accordingly, there is a requirement for up-to-date control techniques to operate effectively the traditional VSIs/CSIs based on VSGs in islanded grids in order to resolve the mentioned problems and deliver a secure power supply to the load side [15-18]. On the other hand, the performance of renewable energy resources and specifically their battery (or storage system) lifespan are vital. Utilizing a VSI as the interface module in the network affects the quality of the resource lifespan negatively because of the lack of providing a continuous input current. Additionally, CSIs based on VSGs require a combination of itself and another in-parallel VSI due to its restricted capability to control the voltage of the grid independently [14,15].

In this matter, using ZSIs is a unique solution to solve most of the above problems in the microgrids for the first time and also in renewable resources (DC power supply). They have more specific characteristics than the traditional inverters to overcome the issues instinctively [19-22]. In this paper, one of the YZSI category [23] has been selected as Fig. 1 (and its equation models in a period) to provide:

- eliminating the switching dead time to solve concerns about short circuits in the inverter, and

consequently increase reliability and decrease distortions like harmonic in output waveforms in addition the Y-source network of the inverter acts as a filter;
- Increasing the inverter efficiency because of the essence of its impedance network, its small size, a single-stage power conversion, soft switching during inductive loads, and sufficient routes for mitigating leakage energy;
- Providing a continuous input current to enhance DC input source lifespan.

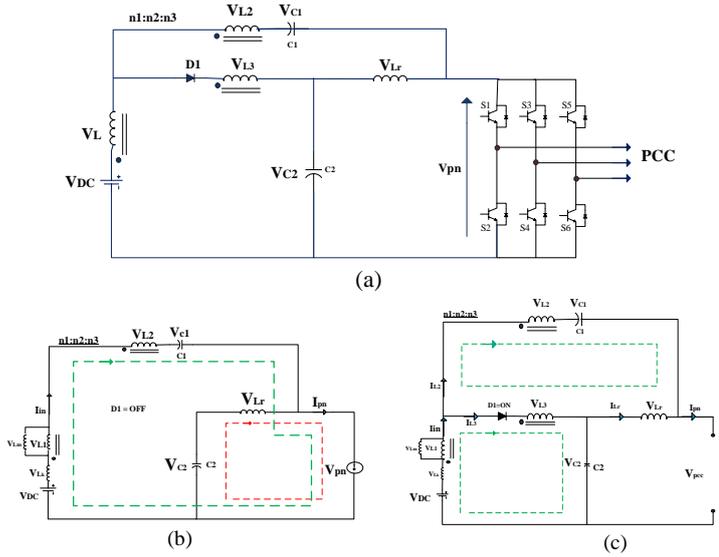

Fig. 1 (a) Configuration of the proposed topology. (b) Equivalent circuit in ST mode. (c) Equivalent Circuit in NST mode.

Equation (1) introduces the AC output voltage with the M modulation coefficient for the inverter and equation (2) is the voltage gain.

$$V_{AC} = \frac{M}{2} \cdot \frac{(1+P)}{(1-d)\cdot(1+P)-d\cdot(1+K)} V_{DC} \quad (1)$$

$$B = \frac{(1+P)}{(1-d)\cdot(1+P)-d\cdot(1+K)} \quad (2)$$

Notably, this paper delivers: (I) The fundamental principles underlying the new enhanced control strategy applied in the proposed YZSI based on VSG construction, along with a comprehensive performance analysis. (II) Accurate analytical findings obtained through rigorous simulations using MATLAB/Simulink and a thorough explanation of the outcomes. (III) Conclusive understandings summarizing the central contributions and findings presented within this paper.

## II. CHARACTERISTICS OF PROPOSED VSG BASED ON YZSI

### A. Introduction of YZSI VSG scheme in islanded Grid

Up to this point, we have observed a great number of configurations involving conventional power electronics converters and various control methodologies founded on VSM. This section presents an innovative synergy between a particular category of Z-source inverters referred to as YZSI and an enhanced version of VSG control tactic. Figure 2 provides a representation of the proposed topology within a microgrid environment featuring distinct load requirements. As depicted in Figure 2, the load requirements are interfaced the electric utility at the Point of Common Coupling (PCC) while being regulated by the newly introduced control framework. When considering the usage of an impedance network converter (ZSI), particularly when integrating it with other systems, it's essential to recognize that these converters are inherently more complex than traditional converters due to their various components. Therefore, it is imperative to cautiously consider several parameters in their design and control. The following section will provide a description of how the proposed YZSI-VSG configuration achieves strong self-regulation.

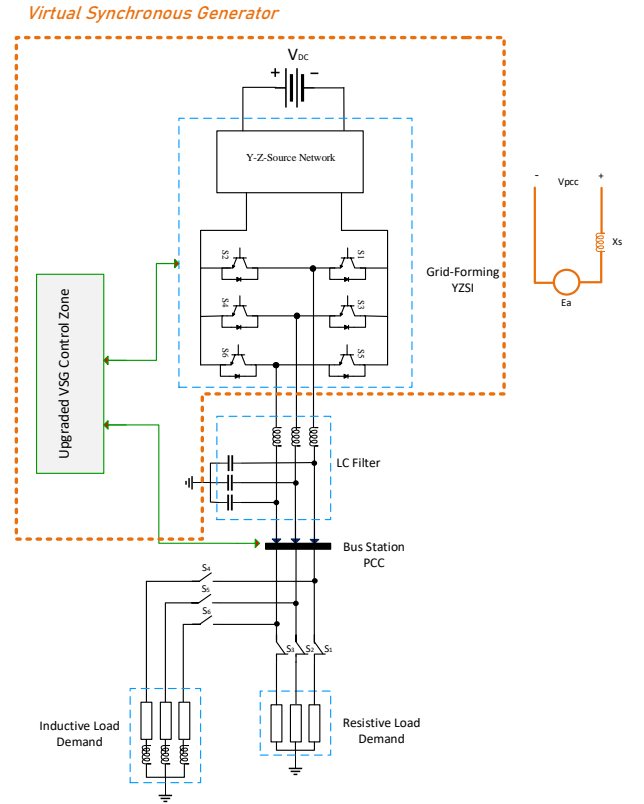

Fig. 2 The proposed islanded grid.

### B. Proposed YZSI-VSG control strategy

Figure 3 illustrates the control method in this paper, which enables concurrent regulation of both the inverter and the grid. The presence of the inverter regulation in the control method enhances overall network security, ensuring a robust end-to-end system. The fundamental elements initiating the operation of this system are the PCC voltage and the inverter voltage. The process commences by comparing the reactive power with the grid's reference value, thereby determining the grid voltage magnitude, symbolized as $V_{com}$. Simultaneously, the control hires a Proportional-Integral (PI) method to compere the reference value and the actual inverter's output voltage to regulate the inverter's current. These two voltage control equations are stated as follows:

$$(Q_{ref} - Q_{meas}).K1 + V_{ref} = |V_{com}| \quad (1)$$

$$(V_{c2\_ref} - V_{c2\_meas}).\left[P1 + I1.\frac{1}{s}\right] \rightarrow \{I_{com}\} \quad (2)$$

In Equation (1), K1 represents the voltage constant, while Method (2) involves P1 and I1 as the proportional and integral gains, respectively. For enhancing the accuracy of the inverter control, it is inspired and valued by calculating the approximate Thevenin equivalent impedance of the Y-impedance network of the inverter in two states (Fig. 1) of a switching period and it has been provided in Equations (3) and (4).

During Shoot Through State (STS)

$$X_L^{Thev.} = \frac{2}{3} X_L \quad (3)$$

During Active State (NSTS)

$$X_L^{Thev.} = \frac{5}{3} X_L \quad (4)$$

$X_L$ is the inductive impedance of the network inductors.

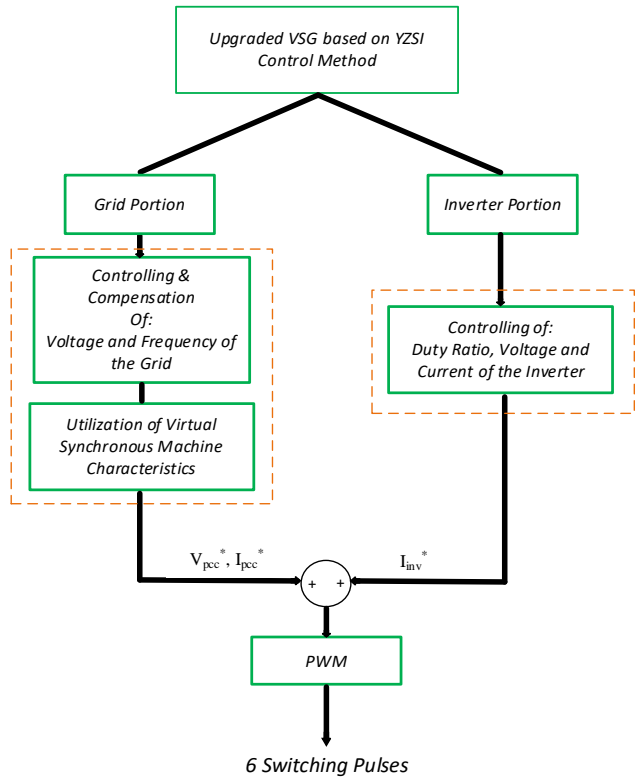

Fig. 3 The algorithm of the proposed control technique.

Equation (5) is focused on the regulation of the grid's frequency. It announces references for active power and frequency, comparing them to their measured values. This value is then passed through a virtual synchronous generator (VSG), which incorporates torque factor (J) and damping coefficient (D) to attain the desired frequency, as demonstrated in Equations (6) and (7), and then following them Equation (8). $K_2$ is the power coefficient, $P_{ref}$ is the basic value of the grid power, and $\theta$ is the voltage phase.

$$(f_{ref} - f_{meas}).K2 + P_{ref} = P_{com} \quad (5)$$

$$\Delta\omega = \frac{1}{J}.\int\left[\frac{(P_{com} - P_{meas})}{2.\pi.f_{ref}} - D.\Delta\omega\right]dt \quad (6)$$

$$\omega_{meas} + \Delta\omega = \omega^* \quad (7)$$

$$\theta = \int \omega^*.dt \quad (8)$$

At the conclusion of this control section, the grid voltage can be computed as per Equation (9) to initialize the inner control loops within the system.

$$|V_{com}|.\sin(\omega^*, \theta) = \{V_a, V_b, V_c\} \quad (9)$$

To boost the precision of the proposed control strategy, an abc/dq0 transformation is employed. This transformation allows us to work exclusively in the dq0 coordinate system, improving control punctuality. Equations (10) and (11) represent the transfer functions for the dq0 and abc coordinate systems, correspondingly.

$$\begin{bmatrix}V_d \\ V_q \\ V_0\end{bmatrix} = \begin{bmatrix}\cos(\theta) & \cos(\theta - \frac{2\pi}{3}) & \cos(\theta + \frac{2\pi}{3}) \\ -\sin(\theta) & -\sin(\theta - \frac{2\pi}{3}) & -\sin(\theta + \frac{2\pi}{3}) \\ \frac{1}{2} & \frac{1}{2} & \frac{1}{2}\end{bmatrix}.\begin{bmatrix}V_a \\ V_b \\ V_c\end{bmatrix} \quad (10)$$

$$\begin{bmatrix}V_a \\ V_b \\ V_c\end{bmatrix} = \left(\begin{bmatrix}\cos(\theta) & \cos\left(\theta - \frac{2\pi}{3}\right) & \cos\left(\theta + \frac{2\pi}{3}\right) \\ -\sin(\theta) & -\sin\left(\theta - \frac{2\pi}{3}\right) & -\sin\left(\theta + \frac{2\pi}{3}\right) \\ \frac{1}{2} & \frac{1}{2} & \frac{1}{2}\end{bmatrix}\right)^{-1}.\begin{bmatrix}V_d \\ V_q \\ V_0\end{bmatrix} \quad (11)$$

In the third stage of the regulator, a double-loop control scheme is employed, as outlined in the following algorithms. This double-loop control is designed to produce the desired grid voltage, grid and inverter currents.

[double-loop control algorithm]:

$$I_{dq0}^* = f(V_{com}^{dq0}, V_{meas}^{dq0}, I_{meas}^{dq0}, I_c)$$

$$V_{dq0}^* = f(I_{dq0}^*, V_{meas}^{dq0}, I_{meas}^{dq0}, V_L)$$

After using equation (11) to inverse the values in dq0 mode to abc mode, the final parameters $V_{pcc}^*$, $I_{pcc}^*$, and $I_{inv}^*$ are produced to employ them in PWM pulse generator.

### III. OPERATIONAL ASSESSMENT

This section consists of two main divisions to expound on the results of the simulation in MATLAB/Simulink. The first part presents the results pertaining to the inverter's performance within the grid. The subsequent part elucidates the grid's behavior under constant load demand, further showcasing the introduction of an inductive load into the system after a delay of 0.4 seconds and then it will be disconnected after 0.8 second.

## A. constituent 1: Evaluation of the Inverter performance

Figure 4 illustrates central waveforms of the inverter and input supply unit. In Fig. 4(a), the DC voltage supply from the battery and the DC link voltage ($V_{pn}$) are presented. Notably, the Y-impedance network operates as a boost converter, achieving a boost factor of 1.2. Fig. 4(b) displays a continuous input current, as previously mentioned, which contributes to prolonging the battery lifespan. Finally, in Fig. 4(c), the produced and delivered active powers from the input supply to the DC link within the proposed virtual system is depicted. The system exhibits an efficiency of 98 percent prior to the introduction of the inductive load, which remains at 97 percent after the joining of the inductive load into the grid. This sustained high efficiency in various scenarios underscores the vigorous and reliable performance of this innovative topology.

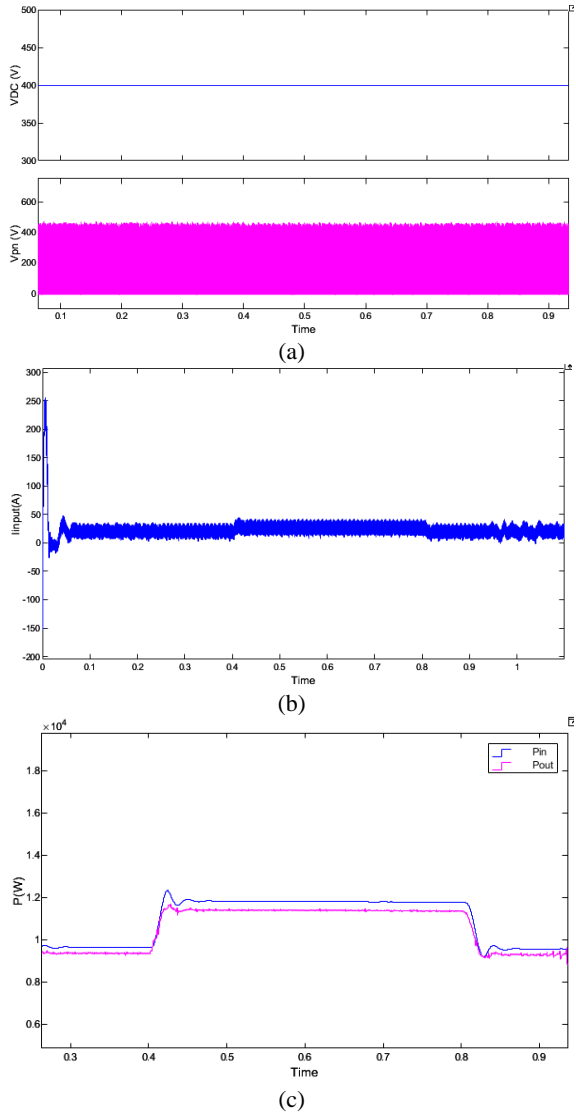

Fig. 4 Performance of the DC supply and the Y-impedance network, a) DC supply and DC link voltage; b) Input current; c) supplied power and power of DC link.

## B. Constituent 2: assessment of the islanded-grid operation utilizing the YZSI-VSG control technique

One of the crucial factors inherent to any power system lies in the quality of the power supplied for customer consumption. In the context of this network, the applied structure guarantees a flawless conveyance of output voltage and current, a fact authenticated by the compelling evidence presented in Figure 5. Accordingly, Fig. 5(a) illustrates the voltage and current profiles within the load demand section across three distinctive scenarios: before the introduction of the inductive load, during its operation, and after its disconnection.

For a more detailed examination, Fig. 5(b) provides a magnified view of the waveform immediately following the entry of the inductive load, while Fig. 5(c) offers an amplified representation of the waveform subsequent to the removal of the inductive load. These visuals serve to shed light on the dynamics of the system during key operational phases. Generally, in an ideal and well-controlled system, it would be aimed for a low THD, typically less than 5% for voltage waveforms, to ensure a seamless and steady power supply Fig. 5(d) proved a small THD in the proposed microgrid. Furthermore, as depicted in Fig. 6(a), the graphical representation of the output power from the supply side underscores its significance which Table 1 illustrates the value of the load demand and another important parameter of the proposed topology in three scenarios. Another serious dimension for assessing quality pertains to the grid frequency. As obvious from Fig. 6(b), the network excellently maintains a constant and reliable frequency across various operational scenarios.

Table 1 and Table 2 collectively deliver information on the central parameters in the network and the control technique, respectively.

Table 1. System Parameters Information.

| Constituent | Quantity |
| --- | --- |
| $V_{DC}$ | 400 V |
| Duty Ratio | 0.2 |
| Switching Frequency | 18 kHz |
| $V_{pcc}$ | 480 V |
| $P_{pcc}$ | 6 & 8 kW |
| $f_{ref}$ | 60 Hz |

Table 2. Control Parameters Information.

| Constituent | Quantity |
| --- | --- |
| $K_1$ | 0.0001 |
| $P_1$ | 100 |
| $I_1$ | 30 |
| D | 20 |
| J | 0.5 |
| $f_{sw}$ | 18 kHz |

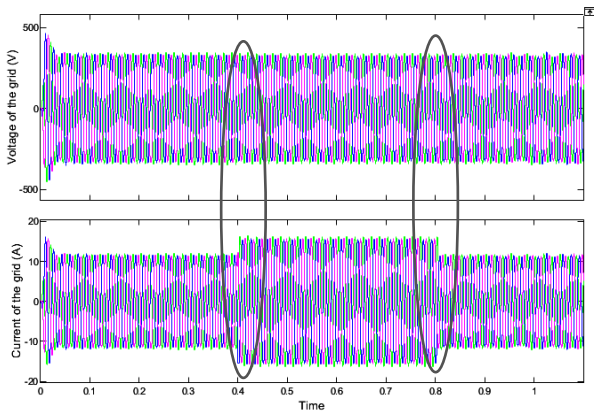

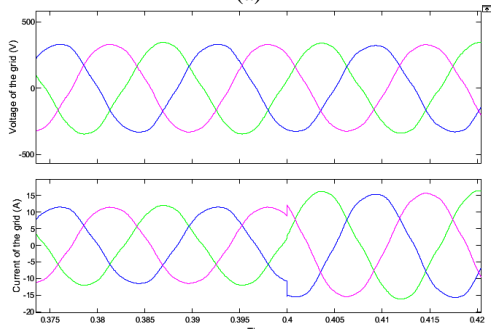

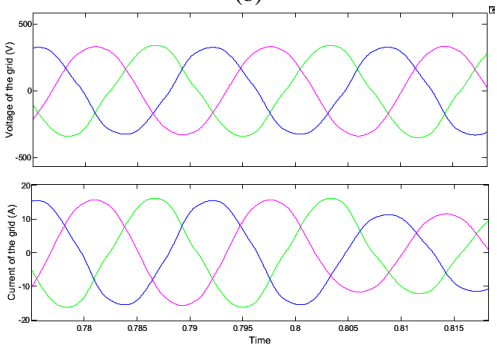

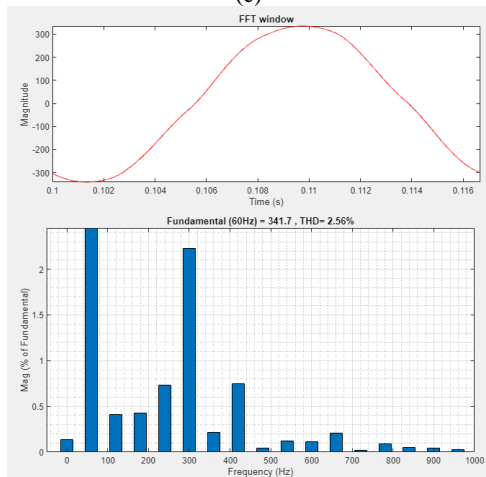

Fig. 5 (a) the voltage and current profiles within the load demand section in three conditions (Before, during, and subsequent to the presence of the inductive load); (b) Zoomed-in waveform after the inductive load entrance; (c) Zoomed-in waveform after the inductive load disconnection; (d) THD of the grid voltage.

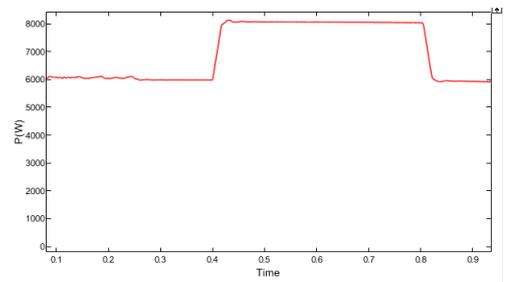

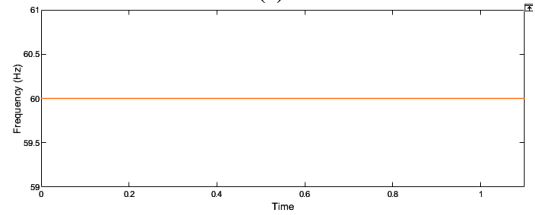

Fig. 6 (a) supplied power to the load demand; (b) and frequency of the grid in the three scenarios.

IV. CONCLUSION

This paper accomplishes with a pioneering achievement in islanded grid applications and renewable energy integration. It introduces a novel fusion of a Y-Source Inverter (YZSI) and an upgraded Virtual Synchronous Generator (VSG) control which not only fulfills the primary control objectives but also it is superbly capable to overcome the instinctive complexities of the utilized impedance source converter. The research emphasizes the critical role of YZSI in addressing the elaborate challenges within islanded grid systems and renewable energy incorporation. It enhances overall system performance, expands power quality, and extends the lifespan of energy resources. The robustness and efficiency of the proposed topology and the control strategy are persuasively authenticated through MATLAB/Simulink simulations. This innovative work embraces significant potential for the future of power systems industry, offering new solutions for contemporary renewable energy configurations.

REFERENCES

[1] Sahoo, S.K., 2016. Renewable and sustainable energy reviews solar photovoltaic energy progress in India: A review. *Renewable and Sustainable Energy Reviews*, *59*, pp.927-939.

[2] Sebestyén, V., 2021. Renewable and Sustainable Energy Reviews: Environmental impact networks of renewable energy power plants. *Renewable and Sustainable Energy Reviews*, *151*, p.111626.

[3] Salehi, M., Darabi, A., Ghaheri, A. and Hoseintabar, M., 2021, May. Design and Analysis of Concentrated Field TFPM Generator for Direct-Drive Wind Turbines. In *2021 29th Iranian Conference on Electrical Engineering (ICEE)* (pp. 335-339). IEEE.

[4] A. Zare, S. D'silva and M. B. Shadmand, "Optimal Ratio of Grid-Forming to Grid-Following Inverters Towards Resilient Power Electronics Dominated Grids," 2023 IEEE Applied Power Electronics Conference and Exposition (APEC), Orlando, FL, USA, 2023, pp. 2347-2352, doi: 10.1109/APEC43580.2023.10131538.

[5] Cheema, K.M., 2020. A comprehensive review of virtual synchronous generator. *International journal of electrical power & energy systems*, *120*, p.106006.

[6] Lu, L. and Cutululis, N.A., 2019, October. Virtual synchronous machine control for wind turbines: a review. In *Journal of Physics: Conference Series* (Vol. 1356, No. 1, p. 012028). IOP Publishing.

[7] S. D'silva, A. Zare, M. B. Shadmand, S. Bayhan and H. Abu-Rub, "Towards Resiliency Enhancement of Network of Grid-Forming and Grid-Following Inverters," in IEEE Transactions on Industrial Electronics, vol. 71, no. 2, pp. 1547-1558, Feb. 2024, doi: 10.1109/TIE.2023.3262866.

[8] Beck, H.P. and Hesse, R., 2007, October. Virtual synchronous machine. In *2007 9th international conference on electrical power quality and utilisation* (pp. 1-6). IEEE.

[9] D'Arco, S. and Suul, J.A., 2013, June. Virtual synchronous machines—Classification of implementations and analysis of equivalence to droop controllers for microgrids. In *2013 IEEE Grenoble Conference* (pp. 1-7). IEEE.

[10] Xue, H. and He, J., 2022, October. A Simplified Power Balance Strategy for Three-Phase Cascaded H-bridge Photovoltaic Inverter. In *2022 IEEE Energy Conversion Congress and Exposition (ECCE)* (pp. 1-6). IEEE.

[11] Modarresi, J., Gholipour, E. and Khodabakhshian, A., 2016. A comprehensive review of the voltage stability indices. *Renewable and Sustainable Energy Reviews*, *63*, pp.1-12.

[12] Cheema, K.M., Chaudhary, N.I., Tahir, M.F., Mehmood, K., Mudassir, M., Kamran, M., Milyani, A.H. and Elbarbary, Z.S., 2022. Virtual synchronous generator: Modifications, stability assessment and future applications. *Energy Reports*, *8*, pp.1704-1717.

[13] Pourghorban, A., Dorothy, M., Shishika, D., Von Moll, A. and Maity, D., 2022, December. Target defense against sequentially arriving intruders. In 2022 IEEE 61st Conference on Decision and Control (CDC) (pp. 6594-6601). IEEE.

[14] Bahrami, M. and Khashroum, Z., 2023. Review of Machine Learning Techniques for Power Electronics Control and Optimization. *arXiv preprint arXiv:2310.04699*.

[15] Liu, J., Miura, Y. and Ise, T., 2015. Comparison of dynamic characteristics between virtual synchronous generator and droop control in inverter-based distributed generators. *IEEE Transactions on Power Electronics*, *31*(5), pp.3600-3611.

[16] Pourghorban, A. and Maity, D., 2023, June. Target defense against a sequentially arriving cooperative intruder team. In *Open Architecture/Open Business Model Net-Centric Systems and Defense Transformation 2023* (Vol. 12544, pp. 65-77). SPIE.

[17] Feleke, S., Pydi, B., Satish, R., Kotb, H., Alenezi, M. and Shouran, M., 2023. Frequency stability enhancement using differential-evolution-and genetic-algorithm-optimized intelligent controllers in multiple virtual synchronous machine systems. *Sustainability*, *15*(18), p.13892.

[18] Pourghorban, A. and Maity, D., 2023. Target defense against periodically arriving intruders. *arXiv preprint arXiv:2303.05577*.

[19] Peng, F.Z., 2003. Z-source inverter. *IEEE Transactions on industry applications*, *39*(2), pp.504-510.

[20] Ellabban, O. and Abu-Rub, H., 2016. Z-source inverter: Topology improvements review. *IEEE Industrial Electronics Magazine*, *10*(1), pp.6-24.

[21] Siwakoti, Y.P., Peng, F.Z., Blaabjerg, F., Loh, P.C. and Town, G.E., 2014. Impedance-source networks for electric power conversion part I: A topological review. *IEEE Transactions on power electronics*, *30*(2), pp.699-716.

[22] Abdelhakim, A., Blaabjerg, F. and Mattavelli, P., 2018. Modulation schemes of the three-phase impedance source inverters—Part II: Comparative assessment. *IEEE Transactions on Industrial Electronics*, *65*(8), pp.6321-6332.

[23] Besati, S., Essakiappan, S. and Manjrekar, M., 2023. A New Flexible Modified Impedance Network Converter. *arXiv preprint arXiv:2304.07866*.